\documentstyle[preprint,aps]{revtex}

\begin{document}

\draft

\preprint{
\begin{tabular}{l}
{\bf HIP-1998-34/TH} \\
\end{tabular} }

\tighten

\title{ Inconsistency of Naive Dimensional Regularizations and Quantum
Correction to Non-Abelian Chern-Simons-Matter Theory Revisited}

\author{M. Chaichian  and W. F. Chen }
 
\address{ High Energy Physics Division, Department of Physics\\
 and\\
 Helsinki Institute of Physics, \\
P.O. Box 9 (Siltavuorenpenger 20 C), FIN-00014 Helsinki, Finland }

\maketitle

\begin{abstract}
We find the inconsistency of dimensional reduction and 
naive dimensional regularization in their applications to Chern-Simons
type gauge theories. Further we adopt a consistent dimensional 
regularization to investigate the quantum correction to non-Abelian 
Chern-Simons term coupled with fermionic matter. Contrary to previous 
results, we find that not only the Chern-Simons coefficient receives 
quantum correction from spinor fields, but the spinor field also gets 
a finite quantum correction.
\end{abstract}

\vspace{3ex}

\section{Introduction}
There has been a considerable amount of popularity in perturbative
Chern-Simons-type theory due to the particular feature of the finite 
renormalization of its coupling constant. Almost all the old
regularization schemes $\cite{ref1,ref2,ref3,ref4}$ and even some 
newly developed ones $\cite{ref5,ref6}$ have been applied. The existence of
the antisymmetric tensor in Chern-Simons term makes the implement
of the regularization method much more non-trivial. Specially the one-loop 
quantum correction to a general three-dimensional field theory 
is very delicate $\cite{ref7}$. Different regularization methods 
can easily produce ambiguity in the quantum corrections.
In particular, it appears that some old regularization schemes can bring 
non-physical quantum corrections$\cite{ref8,ref9}$. 
This makes one be cautious
about the use of some regularization methods.  In this paper we show that 
dimensional reduction and naive dimensional regularization are inconsistent
when they are applied to Chern-Simons type theories. 
Indeed, when we use the consistent
dimensional regularization to re-calculate the one-loop quantum correction
for one typical example, Chern-Simons term coupled with spinor field,
we obtain a result different from the previous ones$\cite{ref10}$. 

Regulating Chern-Simons theory by dimensional reduction means evaluating all
the antisymmetric tensor algebra in three dimensions but
performing the loop momentum integration
in $n$-dimension$\cite{ref10,ref11,ref12}$. The concrete definition
in three dimensions is as follows:
\begin{eqnarray}
&& {\epsilon}^{\mu\nu\rho}{\epsilon}_{\alpha\beta\gamma}
={\delta}^{\mu}_{[~\alpha}{\delta}^{\nu}_{~\beta}\delta^{\rho}_{~\gamma]},~
{\delta}_{~\mu}^{\mu}=3, ~\mu,\nu,\cdots=0,1,2,\nonumber\\
&& \delta^{\mu}_{~\nu}=\tilde{\delta}^{\mu}_{~\nu}+\hat{\delta}^{\mu}_{~\nu},
~\tilde{\delta}^{\mu}_{~\nu}\tilde{\delta}^{\nu}_{~\rho}
=\tilde{\delta}^{\mu}_{~\rho},~
\hat{\delta}^{\mu}_{~\nu}\hat{\delta}^{\nu}_{~\rho}
=\hat{\delta}^{\mu}_{~\rho},
~\tilde{\delta}^{\mu}_{~\nu}\hat{\delta}^{\nu}_{~\rho}=0,\nonumber\\
&&k_{\mu}\tilde{\delta}^{\mu}_{~\nu}=k_{\mu},
~k_{\mu}\hat{\delta}^{\mu}_{~\nu}=0;  ~\tilde{\delta}^{\mu}_{~\mu}=n,
~\hat{\delta}^{\mu}_{~\mu}=3-n.
\label{eq1}
\end{eqnarray}
The inconsistency of this regularization method in four-dimensional
supersymmetric field theories had already 
been found by its inventor$\cite{ref11}$.
For three-dimensional case, it can also be easily shown that this 
regularization method is not consistent. From Eq.(\ref{eq1}), we have
\begin{eqnarray}
\tilde{\epsilon}^{\mu\nu\rho}\tilde{\epsilon}_{\mu\nu\rho}&=&
\tilde{\delta}^{\mu}_{[~\mu}\tilde{\delta}^{\nu}_{~\nu}
\tilde{\delta}^{\rho}_{~\rho]}
=n^3-3n^2+2n=n(n-1) (n-2);\nonumber\\
\hat{\epsilon}^{\mu\nu\rho}\hat{\epsilon}_{\mu\nu\rho}&=&
\hat{\delta}^{\mu}_{[~\mu}\hat{\delta}^{\nu}_{~\nu}\hat{\delta}^{\rho}_{~\rho]}
=(3-n)^3-3(3-n)^2+2(3-n)=(3-n)(2-n)(1-n).
\label{eq2}
\end{eqnarray}
So we can obtain
\begin{eqnarray}
0&=&\left(\tilde{\epsilon}^{\mu\nu\rho}\hat{\epsilon}_{\alpha\beta\gamma}\right)
\left(\tilde{\epsilon}_{\mu\nu\rho}\hat{\epsilon}^{\alpha\beta\gamma}\right)
=\left(\tilde{\epsilon}^{\mu\nu\rho}\tilde{\epsilon}_{\mu\nu\rho}\right)
\left(\hat{\epsilon}_{\alpha\beta\gamma}\hat{\epsilon}^{\alpha\beta\gamma}
\right)\nonumber\\
&=&n(3-n)(n-1)^2(n-2)^2.
\label{eq3}
\end{eqnarray}
Therefore, it is only valid for $n=0,1,2,3$, and thus it is not the analytic 
dimensional continuation as that required by dimensional regularization.

 As for the naive dimensional regularization,
it defines the antisymmetric tensor algebra to satisfy$\cite{ref10}$
\begin{eqnarray}
{\epsilon}_{\mu\sigma\eta}{\epsilon}^{\mu\lambda\tau}
=({\delta}_{~\sigma}^{\lambda}
{\delta}_{~\eta}^{\tau}-{\delta}_{~\sigma}^{\tau}{\delta}_{~\eta}^{\lambda})
{\Gamma}(n-1),
~~{\delta}_{~\sigma}^{\sigma}=n.
\label{eq4}
\end{eqnarray}
We can show that this definition in essence makes the theory defined
in  three dimensions, but not in $n$-dimension as it should be. 
This can be seen from the following simple algebraic manipulations.
 Consider the quantity
${\epsilon}_{\mu\sigma\eta}{\epsilon}^{\mu\lambda\tau}
{\epsilon}_{\alpha\lambda\tau}$. On one hand, it equals to 
\begin{eqnarray}
({\epsilon}_{\mu\sigma\eta}{\epsilon}^{\mu\lambda\tau})
{\epsilon}_{\alpha\lambda\tau}
&=&{\Gamma}(n-1)({\delta}_{~\sigma}^{\lambda}
{\delta}_{~\eta}^{\tau}-{\delta}_{~\sigma}^{\tau}
{\delta}_{~\eta}^{\lambda}){\epsilon}_
{\alpha\lambda\tau}\nonumber\\
&=&{\Gamma}(n-1)({\epsilon}_{\alpha\sigma\eta}
-{\epsilon}_{\alpha\eta\sigma})
=2{\Gamma}(n-1){\epsilon}_{\alpha\sigma\eta};
\label{eq5}
\end{eqnarray}
on the other hand, we have
\begin{eqnarray}
{\epsilon}_{\mu\sigma\eta}({\epsilon}^{\mu\lambda\tau}
{\epsilon}_{\alpha\lambda\tau})
&=&{\epsilon}_{\mu\sigma\eta}{\Gamma}(n-1)
({\delta}_{~\mu}^{\alpha}{\delta}_{~\lambda}
^{\lambda}-{\delta}_{~\lambda}^{\mu}{\delta}_{~\alpha}^{\lambda})\nonumber\\
&=&{\Gamma}(n-1)(n-1){\epsilon}_{\alpha\sigma\eta}.
\label{eq6}
\end{eqnarray}
Comparing Eq.(\ref{eq5}) with Eq.(\ref{eq6}), we can see that
only  $n=3$, otherwise
${\Gamma}(n-1)=0$. Thus the naive dimensional regularization also
does not make the theory well defined.

This motivates us to re-consider Chern-Simons type theory 
in the dimensional regularization  schemes proposed by 't Hooft
and Veltman$\cite{ref13}$. To
our knowledge, up to now this is the only dimensional
continuation scheme compatible with gauge symmetry 
to deal with ${\gamma}_5$ and the similar problems. 
In the following we use this consistent dimensional regularization
to investigate the one-loop quantum corrections of non-Abelian
Chern-Simons term coupled to spinor field, the classical action
 of which in Minkowski space is
\begin{eqnarray}
S={\int}d^3x\left[{\epsilon}^{\mu\nu\rho}
\left(\frac{1}{2}A_{\mu}^a{\partial}_{\nu}A_{\rho}^a
+\frac{1}{3!}gf^{abc}A_{\mu}^aA_{\nu}^bA_{\rho}^c\right)+
\bar{\psi}(i/\hspace{-2mm}{\partial}-m
+g/\hspace{-2.5mm}A^aT^a){\psi}\right],
\label{eq7}
\end{eqnarray}
where ${\psi}$ belongs to the fundamental representation of gauge group, and
for simplicity we only consider the one-flavour case; $\gamma_{\mu}$
($\mu =0,1,2$) are usually chosen as follows$\cite{ref14}$, 
\begin{eqnarray}
&&{\gamma}^0={\sigma}_2,~{\gamma}^1=i{\sigma}_3,~{\gamma}^3=i{\sigma}_1;
\nonumber\\
&&{\gamma}_{\mu}{\gamma}_{\nu}=
g_{\mu\nu}-i{\epsilon}_{\mu\nu\rho}{\gamma}^{\rho}, 
~g_{\mu\nu}=\mbox{diag}(1,-1,-1).
\label{eq8}
\end{eqnarray}
This model has become revived in recent
years owing to its possible physical
application to condensed matter theory. 
The coefficient of Chern-Simons term (called statistical parameter)
plays a crucial role in transmuting the spin and the statistics of the
anyon particles. The quantum corrections up to two-loop for this model was
investigated in dimensional reduction
and naive dimensional regularization$\cite{ref10}$. It was found that there 
exist no quantum corrections and all the renormalization constant
are identically equal to one.

For the case of Chern-Simons typical theory, the dimensional continuation
proposed by 't Hooft and Veltman, which had been been
explicitly written down in Refs.$\cite{ref5,ref18}$, is as follows,
\begin{eqnarray}
&&{\epsilon}^{\mu_1\mu_2\mu_3}{\epsilon}_{\nu_1\nu_2\nu_3}
={\sum}_{{\pi}{\in}P_3}\mbox{sgn}(\pi)\Pi_{i=1}^3
\tilde{\delta}^{\mu_i}_{~\nu_{\pi(i)}}, 
~g_{\mu\nu}=\tilde{g}_{\mu\nu}{\oplus}\hat{g}_{\mu\nu}, 
~p_{\mu}=\tilde{p}_{\mu}{\oplus}\hat{p}_{\mu},\nonumber\\[2mm]
&& {\epsilon}^{\mu\nu\rho}\hat{g}_{\rho\alpha}=0,
 ~ {\epsilon}^{\mu\nu\rho}\hat{p}_{\rho}=0,
~\tilde{\delta}_{~\mu}^{\mu}=3, ~\hat{\delta}_{~\mu}^{\mu}=n-3~.
\end{eqnarray}
Then $n$ is continued to a complex variable  to regulate the theory.
However, the price that must be paid for this consistent definition, as that
pointed out in Ref.$\cite{ref4}$, is first performing 
higher covariant derivative regularization. 
Otherwise this dimensional continuation will lead to linear dependence of
Chern-Simons kinetic operator even after gauge-fixing. As usual, we choose
the simplest higher covariant derivative term, the Yang-Mills Lagrangian, 
\begin{eqnarray}
-\frac{1}{4M}F_{\mu\nu}^aF^{a\mu\nu}, 
~~~F_{\mu\nu}^a={\partial}_{\mu}A_{\nu}^a -
{\partial}_{\nu}A_{\mu}^a+gf^{abc}A_{\mu}^bA_{\nu}^c.
\label{eq10}
\end{eqnarray}
The ghost and gauge-fixing terms have the well known form in the 
covariant gauge,
\begin{eqnarray}
S_{\rm ghost}+S_{\rm g.f.}=\int d^3x \left[-\partial^{\mu}\bar{c}^a
\left(\partial_{\mu}c^a+gf^{abc}A_{\mu}^bc^c\right)
-\frac{1}{2\alpha}(\partial^{\mu}A_{\mu}^a)^2\right].
\end{eqnarray}  
There still exists
another difficulty, that is, this dimensional prescription in fact
defines the $n$-dimensional 
${\epsilon}^{\mu\nu\rho}$  effective  only in  three dimensions
This makes
the regulated theory possess the $SO(3){\otimes}SO(n-3)$ covariance rather 
than $SO(n)$, the regulated propagator will not only take very
complicated form, but it is also not $SO(n)$ covariant. This will make 
the loop integration very difficult to carry out. 
However, thanks to $\cite{ref4}$, 
one can prove that the propagator of gauge
field can be decomposed into 
two parts: one part is composed of evanescent quantity,
which has no contribution to the loop integration in 
the limit $n{\longrightarrow}3$; then one can make use of the 
second part as an effective propagator 
\begin{eqnarray}
G^{ab}_{\mu\nu}(p)=-
\frac{i M}{p^2(p^2-M^2)}
\left(iM{\epsilon}_{\mu\nu\rho}p^{\rho}+p^2g_{\mu\nu}-p_{\mu}p_{\nu}\right)
\label{eq11}
\end{eqnarray}
in order to perform calculation. Note that we have chosen the 
Landau gauge ($\alpha=0$). 
The other Feynman rules are listed as below:

\begin{itemize}
\item Fermion propagator
\begin{eqnarray}
S(p)=i\frac{/\hspace{-2mm}p+m}{p^2-m^2}\delta_{ij};
\label{eq12}
\end{eqnarray}
\item Quark-Gluon vertex
\begin{eqnarray}
ig{\gamma}_{\mu}T^a_{ij}(2\pi)^3 {\delta}^{(3)}(p+q+r).
\label{eq13}
\end{eqnarray}
\end{itemize}

In Sect.II, using the consistent dimensional continuation,
we re-consider some of the new one-loop two-point Green functions 
such as the fermionic self-energy, the ghost self-energy and the
contribution to vacuum polarization tensor from the fermionic loop.
Sect.III is about the one-loop three-point functions such as
the fermion-gluon vertex, the ghost-gluon vertex
and so on. Since it is quite complicated
to straightforwardly calculate the fermion-gluon vertex in a non-Abelian
gauge theory, we make
use of the Slavnov-Taylor identity between fermion-gluon vertex
and the composite ghost-fermion vertex to facilitate the calculation.
In Sect.IV, we define the finite renormalization of the coupling constant
with mass-shell renormalization convention and 
we show that the result is different from that presented 
in the literature previously$\cite{ref3}$. 
Finally, in Sect.V we emphasize our conclusions and discuss the justification
for our results. For clarity and completeness, a derivation of the
needed Slavnov-Taylor identities from BRST symmetry
is presented in Appendix.

\section{One-loop Two-Point Function}

\subsection{Contribution to Vacuum Polarization Tensor from Fermionic Loop}

The contributions to vacuum polarization tensor from the self-interaction of
gauge fields and the ghost loop have been shown in many 
works$\cite{ref3,ref4,ref19}$. Here
we only consider the extra contribution from the spinor 
field. 

The relevant Feynman diagram is shown in Fig.1 and the amplitude is
\begin{eqnarray}
i{\Pi}_{\mu\nu}^{(f)ab}&=&-
g^2\mbox{Tr}(T^aT^b){\int}\frac{d^nk}{(2\pi)^n}
\frac{\mbox{Tr}\left\{{\gamma}_{\nu}[/\hspace{-2mm} k+/\hspace{-2mm} p+m]
{\gamma}_{\mu}
(/\hspace{-2mm} k+m]\right\}}
{(k^2-m^2)[(p+k)^2-m^2]}\nonumber\\[2mm]
&=& -\frac{1}{2}g^2{\delta}^{ab}
{\int}\frac{d^nk}{(2\pi)^n}\frac{-im {\epsilon}_{\mu\nu\rho}p^{\rho}+
2k_{\mu}k_{\nu}+k_{\mu}p_{\nu}+k_{\nu}p_{\mu}-g_{\mu\nu}[k{\cdot}(k+p)-m^2]}
{(k^2-m^2)[(p+k)^2-m^2]},
\label{eq14}
\end{eqnarray}
where we choose the normalization of group factor as,
\begin{eqnarray}
\mbox{Tr}(T^aT^b)=\frac{1}{2}\delta^{ab}.
\end{eqnarray}
Calculation (after taking the limit $n{\longrightarrow}3$) gives
\begin{eqnarray}
i{\Pi}_{\mu\nu}^{(f)ab}&=&\frac{ig^2}{16\pi}{\delta}^{ab}\left\{
i{\epsilon}_{\mu\nu\rho}p^{\rho}\frac{m}{p}
\ln\left[\frac{1+p/(2m)}{1-p/(2m)}\right]
\right.\nonumber\\[2mm]
&-&\left.\left(p^2g_{\mu\nu}-p_{\mu}p_{\nu}\right)\frac{1}{m}
\left[-\frac{m^2}{p^2}+\left(\frac{1}{4}\frac{m}{p}+\frac{m^3}{p^3}\right)
\ln\left(\frac{1+p/(2m)}{1-p/(2m)}\right)\right] \right\},
\label{eq15}
\end{eqnarray}
where $p{\equiv}|p|$.
Using the expansion near $p=0$,
\begin{eqnarray}
\ln\left[\frac{1+p/(2m)}{1-p/(2m)}\right]
=\frac{p}{m}+\frac{1}{12}\frac{p^3}{m^3}
+\frac{1}{80}\frac{p^5}{m^5}+{\cdots},
\label{eq16}
\end{eqnarray}
we have
\begin{eqnarray}
{\Pi}_{\mu\nu}^{(f)ab}(0)=\frac{g^2}{8\pi}{\delta}^{ab}\left[
i{\epsilon}_{\mu\nu\rho}p^{\rho}-\frac{1}{3m}(p^2g_{\mu\nu}-p_{\mu}p_{\nu})
\right]. 
\label{eq17}
\end{eqnarray}
Combining Eq.(\ref{eq15}) with the contributions to polarization tensor 
from gluon and ghost loops$\cite{ref3,ref4,ref19}$, 
\begin{eqnarray}
{\Pi}_{\mu\nu}^{({\rm gl})ab}(p)+{\Pi}_{\mu\nu}^{({\rm gh})ab}(p) 
=-\frac{7}{3}\frac{g^2}{4\pi}C_V\delta^{ab}\epsilon_{\mu\nu\rho}p^{\rho},
\end{eqnarray}
and choosing the renormalization condition on gluon 
mass-shell \footnote{For
pure Chern-Simons theory, this renormalization scheme is
 equivalent to taking the large mass
limit $M{\rightarrow}{\infty}$. However here, owing to the mass parameter
$m$, they are not equivalent.} $p=0$,
\begin{eqnarray}
\Pi^{abR}_{\mu\nu}(0)=0,
\end{eqnarray}
we can define the gluon wave function renormalization constant,
\begin{eqnarray}
Z_A=1-\frac{g^2}{4\pi}\left(\frac{7}{3}C_V+\frac{1}{2}\right).
\end{eqnarray}

\subsection{Self-energy for Spinor Field}

 Let us consider fermionic self-energy. Its Feynman diagram is shown in Fig.2
 and the amplitude is read as
\begin{eqnarray}
-i\Sigma (p,M)&=&-g^2(T^aT^a)M{\int}\frac{d^nk}{(2\pi)^n}
\frac{{\gamma}_{\nu}[/\hspace{-2mm} p+/\hspace{-2mm} k+m]{\gamma}_{\mu}
 [iM{\epsilon}_{\mu\nu\rho} k^{\rho}
+k^2g_{\mu\nu}-k_{\mu}k_{\nu}]}{[(k+p)^2-m^2]k^2(k^2-M^2)}
\nonumber \\[2mm]
&=&-2Mg^2C_2(R){\bf 1} \int\frac{d^nk}{(2\pi)^n}
\frac{(M-/\hspace{-2mm}k)k{\cdot}(k+p)
+m(k^2-\gamma_\alpha k^{\alpha}M)}{[(k+p)^2-m^2]k^2(k^2-M^2)},
\label{eq18}
\end{eqnarray}
where ${\bf 1}$ is the unit matrix in colour space, and we have 
used Eq.(\ref{eq8}) and the identity
\begin{eqnarray}
\int \frac{d^nk}{(2\pi)^n}\frac{{\epsilon}_{\mu\nu\rho} p^{\nu}k^{\rho}}
{[(k+p)^2-m^2]k^2(k^2-M^2)}=0.
\label{eq19} 
\end{eqnarray}
Using the decomposition
\begin{eqnarray}
\frac{1}{k^2(k^2-M^2)}=\frac{1}{M^2}\left(
\frac{1}{k^2-M^2}-\frac{1}{k^2}\right),
\label{eq20}
\end{eqnarray} 
and 
\begin{eqnarray}
k{\cdot}p&=&\frac{1}{2}\{[(k+p)^2-m^2]-k^2-(p^2-m^2)\}\nonumber\\[2mm]
&=&\frac{1}{2}\{[(k+p)^2-m^2]-(k^2-M^2)-(p^2-m^2-M^2)\},
\label{eq21}
\end{eqnarray}
we can write Eq.(\ref{eq18}) as follows: 
\begin{eqnarray}
-i\Sigma (p,M)&=&-2Mg^2C_2(R){\bf 1} \int\frac{d^nk}{(2\pi)^n}\left\{
\left(\frac{p^2-m^2}{2M^2}-\frac{1}{2}-\frac{m}{M}\right)
\frac{/\hspace{-2mm} k-M}{(k^2-M^2) [(k+p)^2-m^2]}\right.\nonumber\\
&+&\frac{1}{2M}\frac{1}{k^2-M^2}
+\frac{p^2-m^2}{2M}\frac{1}{k^2[(k+p)^2-m^2]}\nonumber\\
&-&\left. \left(\frac{p^2-m^2}{2M^2}
-\frac{m}{M}\right)\frac{/\hspace{-2mm}k}{k^2[(k+p)^2-m^2]}\right\}. 
\label{eq22}
\end{eqnarray}
The standard integration gives  
\begin{eqnarray}
-i\Sigma (p,M)&=&-\frac{i}{4\pi}g^2C_2(R){\bf 1}M\left\{1+\frac{p^2-m^2}{Mp}
\ln\left(\frac{1+p/m}{1-p/m}\right)\right.\nonumber\\
&-&\gamma_{\mu}p^{\mu}\left(\frac{p^2-m^2}{2M^2}-\frac{m}{M}
\right)\left[\frac{m}{p^2}+\left(\frac{1}{p}-\frac{m^2}{p^3}\right)
\ln\left(\frac{1+p/m}{1-p/m}\right)\right]\nonumber\\[2mm]
&+&\frac{/\hspace{-2mm}p}{p^2}\left(\frac{p^2-m^2}{2M^2}-\frac{1}{2}
-\frac{m}{M}\right)\left[M-m-\frac{p^2-m^2+M^2}{2p}
\ln\left(\frac{1+(p+m)/M}{1-(p+m)/M}\right) \right]\nonumber\\
&+&\left.\left(\frac{m}{p}+\frac{M}{2p}-\frac{p^2-m^2}{2Mp}\right)
\ln\left(\frac{1+(p+m)/M}{1-(p+m)/M}\right)\right\}.
\label{eq23}
\end{eqnarray}
After taking the large-$M$ limit, we obtain the quark self-energy
\begin{eqnarray}
-i\Sigma (p)&=&-\frac{i}{4\pi}g^2C_2(R){\bf 1}\left\{2M+m
+\frac{p^2-m^2}{p}\ln\left(\frac{1+p/m}{1-p/m}\right)\right.\nonumber\\
&-&\left./\hspace{-2mm}p
\left[\frac{m^2}{p^2}+\left(\frac{m}{p}-\frac{m^3}{p^3}\right)
\ln\left(\frac{1+p/m}{1-p/m}\right)-\frac{2}{3}\right]\right\}.
\label{eq24}
\end{eqnarray}
As usual, this quark self-energy can be written in the form of 
quark mass expansion,
\begin{eqnarray}
\Sigma (p)&=&\frac{1}{2\pi}g^2C_2(R){\bf 1}\left(M+\frac{m}{3}\right)
+\frac{1}{4\pi}g^2C_2(R){\bf 1}\frac{5}{3}(/\hspace{-2mm} p-m)
\nonumber\\
&+&\frac{1}{4\pi}g^2C_2(R){\bf 1}\left\{2m
+\frac{p^2-m^2}{p}\ln\left(\frac{1+p/m}{1-p/m}\right)\right.\nonumber\\
&-&\left./\hspace{-2mm}p
\left[1+\frac{m^2}{p^2}+\left(\frac{m}{p}-\frac{m^3}{p^3}\right)
\ln\left(\frac{1+p/m}{1-p/m}\right)\right]\right\}\nonumber\\
&=&\delta m{\bf 1} -(Z_\psi^{-1}-1)(/\hspace{-2mm}p-m){\bf 1}
 +Z_\psi^{-1}\Sigma_R (p).
\label{eq25}
\end{eqnarray}
Thus in the quark mass-shell renormalization scheme, we have the 
renormalization constants and the radiative correction
of quark self-energy as:
\begin{eqnarray}
m_{\rm ph}&=&m-\delta m=m-\frac{g^2}{2\pi}C_2(R)(M+m);\nonumber\\
Z_\psi&=&1+\frac{5}{3}\frac{g^2}{4\pi}C_2(R);\nonumber\\
\Sigma_R (p)&=&\frac{1}{4\pi}g^2C_2(R){\bf 1}\left\{2m_{\rm ph}
+\frac{p^2-m^2_{\rm ph}}{p}
\ln\left(\frac{1+p/m_{\rm ph}}{1-p/m_{\rm ph}}\right)
\right.\nonumber\\
&-&\left./\hspace{-2mm}p
\left[1+\frac{m^2_{\rm ph}}{p^2}+\left(\frac{m_{\rm ph}}{p}
-\frac{m^3_{\rm ph}}{p^3}\right)
\ln\left(\frac{1+p/m_{\rm ph}}{1-p/m_{\rm ph}}\right)\right]\right\}.
\label{eq26}
\end{eqnarray}

\subsection{Self-energy for Ghost  Field}

The self-energy for ghost field had been explicitly shown in $\cite{ref19}$
in  a different method. Here for completeness and later use, we 
re-calculate it in terms of consistent dimensional regularization (Fig.3),
\begin{eqnarray} 
i\Sigma_g^{(1)ab}(p)p^2&=& 
\lim_{M\to\infty}g^2 C_V\delta^{ab}\int\frac{d^nk}{(2\pi)^n}
\frac{M}{k^2(k^2-m^2) (k+p)^2}\left[k^2p^2-(k.p)^2\right]\nonumber\\
&=&\lim_{M\to\infty}g^2 C_V\delta^{ab}\int\frac{d^nk}{(2\pi)^n}\left[
\frac{Mp^2}{(k^2-M^2)(k+p)^2}-\frac{1}{M}\frac{(k\cdot p)^2}{(k^2-M^2)(k+p)^2}
\right] \nonumber\\
&=&\lim_{M\to\infty}
 g^2C_V\delta^{ab}ip^2\frac{1}{8\pi}\left[\frac{1}{2}+\frac{1}{2}
\frac{M^2}{p^2}-\frac{1}{4} \frac{M^3}{p^3}\left(1-\frac{p^2}{M^2}\right)^2
\ln\left(\frac{1+p/M}{1-p/M}\right)\right]\nonumber\\ 
&=&g^2C_V\delta^{ab}\frac{i}{4\pi}p^2\frac{2}{3}.
\label{eq27}
\end{eqnarray}
Consequently, one can define the wave function renormalization constant
for ghost field,
\begin{eqnarray}
Z_c=1+\frac{g^2}{4\pi}\frac{2}{3}C_V.
\label{eq27n}
\end{eqnarray}

\section{One-loop Three-point function}

\subsection{One-loop On-shell Quantum Correction to Fermion-Gluon Vertex}

Let us see the one-loop quantum correction to quark-gluon vertex, 
which receives contributions from two Feynman diagrams (Fig.4). 
The first diagram is quite simple and can be calculated analytically.
However the calculation for the second digram is quite complicated 
since it contains one three-gluon vertex and two gauge field propagators.
Thus we shall make use of the Slavnov-Taylor identity to convert
the calculation of fermion-gluon vertex into that of composite 
fermion-ghost vertices, whose amplitude can be easily calculated.
The detailed derivation of this identity and its one-loop form
are listed in Appendix. 

From Eq.(\ref{eqa13}), we can see that to calculate the quark-gluon vertex,
three parts need to be considered. The first part is associated
with the ghost field self-energy, which can be easily 
obtained from Eq.(\ref{eq27}),
\begin{eqnarray} 
\Sigma_g^{(1)}(p)=\frac{g^2}{4\pi}\frac{2}{3}C_V.
\label{eq28}
\end{eqnarray}

Now we turn to the second part, which is connected with the quark self-energy.
To calculate its contribution to the on-shell quark-gluon vertex, we should
first pull out the factor $(q-p)_{\mu}$ and then put it on mass-shell. 
From Eq.(\ref{eq24}), we have 
\begin{eqnarray}
&&-i\left[\Sigma^{(1)}(q)-\Sigma^{(1)}(p)\right]\nonumber\\
&& =-2g^2C_2(R)M\left\{\left(m+\frac{M}{2}\right)\int\frac{d^3k}{(2\pi)^3}
\frac{1}{k^2-M^2}\left[\frac{1}{(k+q)^2-m^2}
-\frac{1}{(k+p)^2-m^2}\right]\right.\nonumber\\
&-&\left(\frac{m}{M}+\frac{1}{2}\right)\int\frac{d^nk}{(2\pi)^n}
\frac{1}{(k^2-M^2)}\left[\frac{/\hspace{-2mm}k}{(k+q)^2-m^2}
-\frac{/\hspace{-2mm}k}{(k+p)^2-m^2}\right]\nonumber\\
&+&\frac{m}{M}\int\frac{d^nk}{(2\pi)^n}
\left[\frac{/\hspace{-2mm}k}{k^2\left[(k+q)^2-m^2\right]}
-\frac{/\hspace{-2mm}k}{k^2\left[(k+p)^2-m^2\right]}\right]\nonumber\\
&-&\frac{q^2-m^2}{2M}\int\frac{d^3k}{(2\pi)^3}
\frac{1}{(k^2-M^2)\left[(k+q)^2-m^2\right]}
+\frac{p^2-m^2}{2M}\int\frac{d^3k}{(2\pi)^3}
\frac{1}{(k^2-M^2)\left[(k+p)^2-m^2\right]} \nonumber\\ 
&+&\frac{q^2-m^2}{2M^2}\int\frac{d^3k}{(2\pi)^3}
\frac{/\hspace{-2mm}k}{(k^2-M^2)\left[(k+q)^2-m^2\right]}
-\frac{p^2-m^2}{2M^2}\int\frac{d^3k}{(2\pi)^3}
\frac{/\hspace{-2mm}k}{(k^2-M^2)\left[(k+p)^2-m^2\right]}
\nonumber\\
&+&\left.\frac{q^2-m^2}{2M}\int\frac{d^3k}{(2\pi)^3}
\frac{1}{k^2\left[(k+q)^2-m^2\right]}
-\frac{p^2-m^2}{2M}\int\frac{d^3k}{(2\pi)^3}
\frac{1}{k^2\left[(k+p)^2-m^2\right]}\right\} \nonumber\\
&=&2(q-p)^{\mu}g^2C_2(R)M\left\{\left(m+\frac{M}{2}\right)
\int\frac{d^3k}{(2\pi)^3}\frac{2k_{\mu}}{(k^2-M^2)\left[(k+q)^2-m^2\right]
\left[(k+p)^2-m^2\right]}\right.\nonumber\\
&-&\left(\frac{m}{M}+\frac{1}{2}\right)\int\frac{d^3k}{(2\pi)^3}
\frac{2/\hspace{-2mm}kk_{\mu}}{(k^2-M^2)\left[(k+q)^2-m^2\right]
\left[(k+p)^2-m^2\right]}\nonumber\\
&+&\frac{m}{M}\int\frac{d^3k}{(2\pi)^3}
\frac{2/\hspace{-2mm}kk_{\mu}}{k^2\left[(k+q)^2-m^2\right]
\left[(k+p)^2-m^2\right]} \nonumber\\
&-&\frac{q_{\mu}+p_{\mu}}{2M}\int\frac{d^3k}{(2\pi)^3}
\frac{M^2}{k^2(k^2-M^2)\left[(k+q)^2-m^2\right]}\nonumber\\
&-&\frac{q_{\mu}+p_{\mu}}{2M^2}\int\frac{d^3k}{(2\pi)^3}
\frac{/\hspace{-2mm}k}{(k^2-M^2)\left[(k+q)^2-m^2\right]}\nonumber\\
&-&\frac{p^2-m^2}{2M}
\int\frac{d^3k}{(2\pi)^3}\frac{2M^2k_{\mu}}{k^2 (k^2-M^2) 
\left[(k+q)^2-m^2\right]
\left[(k+p)^2-m^2\right]}\nonumber\\ 
&+&\left.\frac{p^2-m^2}{2M^2}\int\frac{d^3k}{(2\pi)^3}
\frac{2/\hspace{-2mm}kk_{\mu}}{(k^2-M^2)\left[(k+q)^2-m^2\right]
\left[(k+p)^2-m^2\right]} \right\},
\label{eq29}
\end{eqnarray}
where we have thrown away the vanishing terms in the large-$M$ limit. 

As above, to compute the terms in Eq.(\ref{eq29}),
we cannot take the large-$M$ limit directly. So we still make use of
above decomposition, and then we have 
\begin{eqnarray}
&&-i\left[\Sigma^{(1)}(q)-\Sigma^{(1)}(p)\right]
= 2(q-p)^{\mu}g^2C_2(R)\left\{-\int\frac{d^nk}{(2\pi)^n}
\frac{k_{\mu}}{(k^2+2k{\cdot}p)(k^2+2k{\cdot}q)}\right.\nonumber\\
&-&\left(m+\frac{M}{2}\right)
\int\frac{d^nk}{(2\pi)^n}\frac{/\hspace{-2mm}kk_{\mu}}{(k^2-M^2)(k^2-m^2)^2}
+m\int\frac{d^nk}{(2\pi)^n}
\frac{/\hspace{-2mm}kk_{\mu}}{k^2(k^2+2k{\cdot}p)(k^2+2k{\cdot}q)}
\nonumber\\
&-&\frac{q_{\mu}+p_{\mu}}{2M}\int \frac{d^nk}{(2\pi)^n}
\frac{/\hspace{-2mm}k}{(k^2-M^2)(k^2+2k{\cdot}q)}
+\left.\frac{p^2-m^2}{M}\int \frac{d^nk}{(2\pi)^n}
\frac{/\hspace{-2mm}kk_{\mu}}{(k^2-M^2)(k^2-m^2)^2}
\right\}\nonumber\\
&=&2(q-p)^{\mu}g^2C_2(R)\left\{\frac{i}{16\pi}(p_{\mu}+q_{\mu})
\frac{1}{r}\ln\left(\frac{1+r/(2m)}{1-r/(2m)}\right)+
\frac{i}{4\pi}\frac{\gamma_{\mu}}{3}\right.\nonumber\\
&&+\left.
\frac{i}{8\pi}\left[\frac{m}{r}\ln\left(\frac{1+r/(2m)}{1-r/(2m)}\right)
\gamma_{\mu}
-\frac{p_{\mu}+q_{\mu}}{m}\frac{2}{4-r^2/m^2}\right]\right\}\nonumber\\
&=&(q-p)^{\mu}\frac{i}{2\pi}g^2C_2(R)\left\{\frac{p_{\mu}+q_{\mu}}{m}\left[
\frac{m}{4r}\ln\left(\frac{1+r/(2m)}{1-r/(2m)}\right)-
\frac{1}{4-r^2/m^2}\right]\right.\nonumber\\
&&\left.+\gamma_{\mu}\left[\frac{1}{3}
+\frac{m}{2r}\ln\left(\frac{1+r/(2m)}{1-r/(2m)}\right)
\right] \right\},
\label{eq30} 
\end{eqnarray}
where $r_{\mu}=q_{\mu}-p_{\mu}$. 
   
 Now we consider the contribution from the one-loop on-shell 
composite ghost-fermion vertex. The corresponding Feynman diagrams are
shown in Fig.5 and the amplitude we need reads as follows:
\begin{eqnarray}
i\gamma^{(1)} &{\equiv}&
g^2\bar{u}(q)\left[\gamma^{(1)a}(p,q,r)(/\hspace{-2mm}q-m)-
(/\hspace{-2mm}p-m)\gamma^{(1)a}(p,q,r)\right]u(p)\nonumber\\
&=&(q-p)^{\mu}\frac{1}{4}g^2C_VT^a \bar{u}(q)\left[\int\frac{d^nk}{(2\pi)^n}
\frac{(/\hspace{-2mm}k+/\hspace{-2mm}p+m)(\gamma_{\mu}/\hspace{-2mm}k
-/\hspace{-2mm}k\gamma_{\mu})}{k^2\left[(k+p)^2-m^2\right]
\left[k-r\right]^2}(/\hspace{-2mm}q-m)\right.\nonumber\\
&& +\left.\int\frac{d^nk}{(2\pi)^n}(/\hspace{-2mm}p-m)
\frac{(/\hspace{-2mm}k\gamma_{\mu}-\gamma_{\mu}/\hspace{-2mm}k)(/\hspace{-2mm}k
+/\hspace{-2mm}q+m)}{(l+r)^2l^2\left[(l+p)^2-m^2\right]}\right]u(p)\nonumber\\
&=&(q-p)^{\mu}\frac{1}{2}g^2C_VT^a \bar{u}(q)\left\{\int\frac{d^nk}{(2\pi)^n}
\left[\frac{/\hspace{-2mm}kk_{\mu}
+(/\hspace{-2mm}p+m)(k_{\mu}-/\hspace{-2mm}k\gamma_{\mu})}
{k^2\left[(k+p)^2-m^2\right](l-r)^2}\right.\right.\nonumber\\
&& \left.-\frac{\gamma_{\mu}}{\left[(k+p)^2-m^2\right]
(l-r)^2}\right](/\hspace{-2mm}q-m)\nonumber\\
&&+\left.\int\frac{d^nk}{(2\pi)^n}(/\hspace{-2mm}p-m)
\left[\frac{/\hspace{-2mm}kk_{\mu}+(k_{\mu}-\gamma_{\mu}/\hspace{-2mm}k)
(\gamma_{\mu}/\hspace{-2mm}q+m)}{(l+r)^2l^2\left[(l+q)^2-m^2\right]}-
\frac{\gamma_{\mu}}{\left[(k+q)^2-m^2\right]
(l+r)^2}\right]\right\}u(p)\nonumber\\
&=&(q-p)^{\mu}\left[\gamma^{(1)}_{\mu}+\gamma^{(2)}_{\mu}
+\gamma^{(3)}_{\mu}+\gamma^{(4)}_{\mu}\right],
\label{eq31}
\end{eqnarray}
where we have taken the large-$M$ limit and used the mass-shell condition. 
Correspondingly, we have
\begin{eqnarray}
\gamma^{(1)}_{\mu}&=&-\frac{g^2C_VT^a}{2}\int\frac{d^nk}{(2\pi)^n}
\left[\frac{2(q_{\mu}-m\gamma_{\mu})}{(k^2+2k\cdot p)(k-r)^2}
+\frac{2(p_{\mu}-m\gamma_{\mu})}{(k^2+2k\cdot q)(k+r)^2}\right]\nonumber\\
&=&g^2C_VT^a\frac{i}{8\pi}\left[(m\gamma_{\mu}-q_{\mu})\frac{1}{q}
\ln\left(\frac{1+q/m}{1-q/m}\right)|_{q^2=m^2}\right.\nonumber\\
&&\left.+(m\gamma_{\mu}-p_{\mu})\frac{1}{p}
\ln\left(\frac{1+p/m}{1-p/m}\right)|_{p^2=m^2}\right]\nonumber\\
&=&g^2C_VT^a\frac{i}{8\pi}\left[\frac{p_{\mu}+q_{\mu}}{m}-2\gamma_{\mu}\right]
\left(\frac{1}{\epsilon_{\rm IR}}+\ln\frac{\mu}{m}\right);
\label{eq32}
\end{eqnarray}
\begin{eqnarray}
\gamma^{(2)}_{\mu}&=&\frac{g^2C_VT^a}{2}\int\frac{d^nk}{(2\pi)^n}
\left[\frac{2q{\cdot}kk_{\mu}}{k^2(k^2+2p\cdot k)(k-r)^2}
+\frac{2p{\cdot}kk_{\mu}}{k^2(k^2+2q\cdot k)(k+r)^2}\right.\nonumber\\
&&+\left.\frac{k_{\mu}(2p\cdot q-2 m^2)}{k^2(k^2+2k\cdot p)(k-r)^2}+
\frac{k_{\mu}(2p\cdot q-2 m^2)}{k^2(k^2+2k\cdot q)(k+r)^2}\right]\nonumber\\
&=&\frac{g^2C_VT^a}{2}\int\frac{d^nk}{(2\pi)^n}\left[\frac{k_{\mu}}{k^2(k-r)^2}
-\frac{k_{\mu}}{k^2(k^2+2k\cdot p)}+\frac{r^2k_{\mu}}{k^2(k^2+2k\cdot p)(k-r)^2}\right.
\nonumber\\
&&+\frac{k_{\mu}}{k^2(k+r)^2}
-\frac{k_{\mu}}{k^2(k^2+2k\cdot q)}+\frac{r^2k_{\mu}}{k^2(k^2+2k\cdot q)(k+r)^2}\nonumber\\
&&+\left.\frac{k_{\mu}(2p\cdot q-2 m^2)}{k^2(k^2+2k\cdot p)(k-r)^2}+
\frac{k_{\mu}(2p\cdot q-2 m^2)}{k^2(k^2+2k\cdot q)(k+r)^2}\right]\nonumber\\
&=&-\frac{g^2C_VT^a}{2}\int\frac{d^nk}{(2\pi)^n}\left[\frac{k_{\mu}}{k^2(k^2+2k\cdot p)}
+\frac{k_{\mu}}{k^2(k^2+2k\cdot q)}\right]\nonumber\\
&=&\frac{i}{16\pi}C_VT^ag^2\frac{p_{\mu}+q_{\mu}}{m};
\label{eq33}
\end{eqnarray}
\begin{eqnarray}
\gamma^{(3)}_{\mu}&=&g^2C_VT^a m\int\frac{d^nk}{(2\pi)^n}\left[
\frac{/\hspace{-2mm}kk_{\mu}}{k^2(k^2+2p\cdot k)(k-r)^2}
+\frac{/\hspace{-2mm}kk_{\mu}}{k^2(k^2+2q\cdot k)(k+r)^2}\right]|_{p^2=q^2=m^2}
\nonumber\\
&=&g^2C_VT^a\frac{i}{32\pi}\left[m\gamma_{\mu}-\frac{p_{\mu}+q_{\mu}}{2}\right]
\left[\frac{1}{q}
\ln\left(\frac{1+q/m}{1-q/m}\right)|_{q^2=m^2}\right.\nonumber\\
&&\left.+
\frac{1}{p}\ln\left(\frac{1+p/m}{1-p/m}\right)|_{p^2=m^2}\right]
+\frac{i}{16\pi}\left(\gamma_{\mu}-\frac{3(p_{\mu}+q_{\mu})}{2}\right)
\nonumber\\
&=&-g^2C_VT^a\frac{i}{16\pi}\left[\frac{p_{\mu}+q_{\mu}}{2m}-\gamma_{\mu}\right]
\left(\frac{1}{\epsilon_{\rm IR}}+\ln\frac{\mu}{m}\right)
+\frac{i}{16\pi}\left[\gamma_{\mu}-\frac{3(p_{\mu}+q_{\mu})}{2}\right],
\label{eq33i}
\end{eqnarray}
where $\epsilon_{\rm IR}=3-n$ and $\mu$ is
the artificial parameter with mass dimension. One notices that 
there is an IR pole term, which is induced purely
by the mass-shell condition.
During the above procedure, we have used the identities
\begin{eqnarray}
2 p\cdot k&=&(k^2+2 p\cdot k)-k^2,\nonumber\\
2k\cdot r &=& (k+r)^2-k^2-r^2=k^2+r^2-(k-r)^2,
\label{eq34}
\end{eqnarray}
and the mass-shell condition
\begin{eqnarray}
2p\cdot q-2m^2=p^2+q^2-(q-p)^2-2m^2=-r^2.
\label{eq35}
\end{eqnarray}

As for $\gamma^{(4)}_{\mu}$, it is a little more complicated, namely
\begin{eqnarray}
\gamma^{(4)}_{\mu}&=&-\frac{1}{2}g^2C_VT^a\int\frac{d^nk}{(2\pi)^n}
\left[ \frac{4p\cdot k q_{\mu}-4q\cdot k p_{\mu}+2(p\cdot q-m^2)/\hspace{-2mm}k
\gamma_{\mu} }{k^2(k^2+2k\cdot p)(l-r)^2}\right.\nonumber\\
&&+\frac{4mp_{\mu}/\hspace{-2mm}k-4mk\cdot p\gamma_{\mu}}
{k^2(k^2+2k\cdot p)(l-r)^2}
+\frac{4q\cdot kp_{\mu}-4k\cdot p q_{\mu}+2(p\cdot q-m^2)\gamma_{\mu}
/\hspace{-2mm}k}{k^2 (k^2+2k\cdot q)(l+r)^2}\nonumber\\
&&+\left.\frac{4mq_{\mu}/\hspace{-2mm}k-4mk\cdot q\gamma_{\mu}}
{k^2 (k^2+2k\cdot q)(l+r)^2}\right].
\label{eq36}
\end{eqnarray}
However, from Eq.(\ref{eq36}) one can see that 
we only need two Feynman integrals (on-shell),
\begin{eqnarray}
\int\frac{d^nk}{(2\pi)^n}\frac{k_{\mu}}{k^2(k^2+2k\cdot p)(k-r)^2}|_{p^2=m^2}&=&
Ap_{\mu}+Br_{\mu};
\nonumber\\
\int\frac{d^nk}{(2\pi)^n}\frac{k_{\mu}}{k^2(k^2+2k\cdot q)(k+r)^2}|_{q^2=m^2}
&=&Aq_{\mu}-Br_{\mu},
\label{eq37}
\end{eqnarray}
where $A$ and $B$ are the form factors needed to be determined.  
It is easy to obtain that
\begin{eqnarray}
A&=&\frac{1}{2(p^2-q^2)}\left\{\int\frac{d^nk}{(2\pi)^n}\left[
\frac{1}{\left[(k+p)^2-m^2\right](k+r)^2}
-\frac{1}{\left[(k+q)^2-m^2\right](k+r)^2}\right]|_{p^2=q^2=m^2}\right.
\nonumber\\
&&+\left.\int\frac{d^nk}{(2\pi)^n}\left[\frac{q^2-m^2}
{k^2\left[(k+q)^2-m^2\right](k+r)^2}
- \frac{p^2-m^2}
{k^2\left[(k+p)^2-m^2\right](k-r)^2}\right]|_{p^2=q^2=m^2}\right\}
\nonumber\\
&=&\frac{i}{16\pi}\frac{1}{m(p^2-q^2)}\left[\frac{1}{q}
\ln\left(\frac{1+q/m}{1-q/m}\right)
-\frac{1}{p}\ln\left(\frac{1+p/m}{1-p/m}\right)\right]|_{p^2=q^2=m^2} 
-\frac{I}{2}\nonumber\\
&=&\frac{i}{16\pi}\frac{1}{(p^2-q^2)m^{1+\epsilon}}\int^1_0 dx \left[
\frac{1}{\left[x-q^2/m^2 x (1-x)\right]^{(1+\epsilon)/2} }|_{p^2=m^2}
\right.  \nonumber\\
&&\left.-\frac{1}{\left[x-p^2/m^2 x (1-x)\right]^{(1+\epsilon)/2}}
\right]|_{q^2=m^2}-\frac{I}{2}
\nonumber\\
&=&\frac{i}{16\pi} \frac{1+\epsilon}{2}\frac{1}{m^{3+\epsilon}}
\int^1_0 dx x (1-x)\nonumber\\
&&\times \int_0^1 dy\frac{1}{\left\{\left[x-q^2/m^2 x (1-x)\right]
+(q^2-p^2)/m^2 x (1-x) y\right\}^{(3+\epsilon)/2}}|_{p^2=q^2=m^2}-\frac{I}{2}\nonumber\\
&=&\frac{i}{16\pi} \frac{1+\epsilon}{2}\frac{1}{m^{3+\epsilon}}
\int^1_0 dx (1-x) x^{-2-\epsilon}-\frac{I}{2}
=\frac{i}{32\pi}\frac{1}{m^3}\left(\frac{1}{\epsilon_{IR}}
+\ln\frac{\mu}{m}\right)-\frac{I}{2},
\label{eq38}
\end{eqnarray}
where we have used the relation
\begin{eqnarray}
\frac{1}{a^{(1+\epsilon)/2}}-\frac{1}{b^{(1+\epsilon)/2}}
=\frac{1+\epsilon}{2}\int_0^1 dy \frac{b-a}{\left[a+(b-a)y
\right]^{(3+\epsilon)/2}},
\label{eq39}
\end{eqnarray}
and 
\begin{eqnarray}
I&{\equiv}&\int\frac{d^nk}{(2\pi)^n} \frac{1}
{k^2\left[(k+q)^2-m^2\right](k+r)^2}|_{q^2=m^2}\nonumber\\
&=&\int\frac{d^nk}{(2\pi)^n} \frac{1}
{k^2\left[(k+p)^2-m^2\right](k-r)^2}|_{p^2=m^2}\nonumber\\
&=& \frac{i}{4\pi^2}\Gamma\left(\frac{3-\epsilon}{2}\right)
\Gamma\left(\frac{3+\epsilon}{2}\right)\frac{1}{m^3}
\left(\frac{\mu}{m}\right)^{\epsilon}\nonumber\\
&\times& \int_0^1dx\int_0^1dy
\frac{y}{\left[(1-x)^2y^2-k^2/m^2 x y (1-y)\right]^{(3+\epsilon)/2}}.
\end{eqnarray}
Thus we obtain
\begin{eqnarray}
\gamma^{(4)}_{\mu}&=&-\frac{1}{2}g^2C_VT^aA\left[(4m^2-2 r^2)(p_{\mu}+q_{\mu})+
(2mr^2-8m^3)\gamma_{\mu}\right]\nonumber\\
&=&-\frac{i}{32\pi}g^2C_VT^a\left(\frac{1}{\epsilon_{\rm IR}}+
\ln\frac{\mu}{m}-16m^3I\right)
 \left[\left(2- \frac{r^2}{m^2}\right)\frac{p_{\mu}+q_{\mu}}{m}+
\left(\frac{r^2}{m^2}-4\right)\gamma_{\mu}\right].
\label{eq40}
\end{eqnarray}
and hence
\begin{eqnarray}
i\gamma^{(1)}&=&(q-p)^{\mu}g^2C_VT^a\frac{i}{8\pi}
\left\{\frac{p_{\mu}+q_{\mu}}{m}\left[
\frac{1}{4}\left(\frac{1}{\epsilon_{\rm IR}}
\ln\frac{\mu}{m}\right)\left(1+\frac{r^2}{m^2}\right)
-\frac{1}{4}+4m^3I\left(2-\frac{r^2}{m^2}\right)\right]\right.\nonumber\\
&&\left. -\gamma_{\mu}\left[\frac{1}{2}\left(\frac{1}{\epsilon_{\rm IR}}
+\ln\frac{\mu}{m}\right)\left(1+\frac{r^2}{2m^2}\right)-\frac{1}{2}
+4m^3I\left(\frac{r^2}{m^2}-4\right)\right]\right\}.
\label{eq42}
\end{eqnarray}

From Eqs.(\ref{eqa13}), (\ref{eq28}), (\ref{eq30}) and (\ref{eq42}), we 
finally obtain the on-shell one-loop fermion-gluon vertex,
\begin{eqnarray}
\Gamma_{\mu}^{(1)a}(r)&=&-\frac{g^2}{4\pi}T^a\left\{\gamma_{\mu}
\left[-C_V\left(\frac{5}{12}
+\frac{1}{4}\left(\frac{1}{\epsilon_{\rm IR}}+\ln\frac{\mu}{m}\right)
\left(1+\frac{r^2}{2m^2}\right)\right.\right.\right.\nonumber\\
&& \left.\left. -2m^3I\left(\frac{r^2}{m^2}-4\right)\right)
+C_2(R)\left(\frac{2}{3}+\frac{m}{r}
\ln\frac{1+r/(2m)}{1-r/(2m)}\right)\right]
\nonumber\\
&&+\frac{p_{\mu}+q_{\mu}}{m}\left[C_2(R)
\left(\frac{m}{2r}\ln\left(\frac{1+r/(2m)}{1-r/(2m)}\right)
-\frac{2}{4-r^2/m^2}
-\frac{1}{2}\left(\frac{1}{\epsilon_{\rm IR}}+\ln\frac{\mu}{m}\right)\right)
\right. \nonumber\\
&&\left.\left.+C_V\left(
\frac{1}{8}\left(\frac{1}{\epsilon_{\rm IR}}+\ln\frac{\mu}{m}\right)
 \left(1+\frac{r^2}{m^2}\right)
-\frac{1}{8}+2m^3I\left(2-\frac{r^2}{m^2}\right)\right)\right]\right\}
\nonumber\\
&=&-\frac{g^2}{4\pi}T^a\gamma_{\mu}\left\{C_V\left[-\frac{2}{3}+
\frac{1}{8}\left(\frac{1}{\epsilon_{\rm IR}}+\ln\frac{\mu}{m}-16m^3I\right)
\frac{r^2}{m^2}\right]\right.
\nonumber\\
&&\left.+C_2(R)\left[\frac{2}{3}
+\frac{3}{2}\frac{m}{r}\ln\left(\frac{1+r/(2m)}{1-r/(2m)}\right)
-\frac{2}{4-r^2/m^2} \right]\right\}\nonumber\\
&&+\frac{g^2}{4\pi}T^a\frac{i\epsilon_{\mu\nu\rho}r^{\nu}\gamma^{\rho}}{m} 
\left\{C_2(R)
\left[\frac{m}{2r}\ln\left(\frac{1+r/(2m)}{1-r/(2m)}\right)-\frac{2}{4-r^2/m^2}
\right]\right.\nonumber\\
&&\left.
+C_V\left[\frac{1}{8}\left(\frac{1}{\epsilon_{\rm IR}}+\ln\frac{\mu}{m}\right)
\left(1+\frac{r^2}{m^2}\right)
-\frac{1}{8}+2m^3I\left(2-\frac{r^2}{m^2}\right)\right]\right\}\nonumber\\
&=&\gamma_{\mu}T^af_1(r)+iT^a\epsilon_{\mu\nu\rho}r^{\nu}\gamma^{\rho}f_2(r),
\label{eq43}
\end{eqnarray}
where we have used the three-dimensional analogue of the Gordon identity:
\begin{eqnarray}
{\gamma}_{\mu}=\frac{1}{2m}\left[(p_{\mu}+q_{\mu})
+i{\epsilon}_{\mu\nu\lambda}r^{\nu}
{\gamma}^{\lambda}\right].
\label{eq44}
\end{eqnarray}
We define the vertex at the renormalization point $r=0$ as that done in 
$\cite{ref20}$,
\begin{eqnarray}
&&\Gamma_{\mu}^{a(R)}(0)=0;\nonumber\\
&& \Gamma^{a(R)}_{\mu}(r)=T^a{\gamma}_{\mu}(Z_3^{-1}-1)+Z_3^{-1}
{\Gamma}^{a(R)}_{\mu}(r),
\label{eq45}
\end{eqnarray}
then the quark-gluon vertex renormalization constant is
\begin{eqnarray}
Z_{3}^{-1}&=&1+f_1(0),
\nonumber\\[2mm]
Z_3&=&1+\frac{g^2}{4\pi}\left[-\frac{2}{3}C_V+C_2(R)\frac{5}{3}\right].
\label{eq46}
\end{eqnarray}
One can notice that actually $f_2(0)$ does not vanish. This will induce
a non-minimal (colour) magnetic moment interaction between two 
three-dimensional quarks, which may play a certain
role in the application of this model to condensed matter physics. The similar
result had also been obtained in Abelian case$\cite{ref21}$.  

\subsection{ One-loop Ghost-Gluon Vertex}

For discussing the renormalization of coupling constant, we shall
have a brief look at the one-loop ghost-gluon vertex (Fig.6), 
whose value had been predicted in Ref.$\cite{ref18}$ 
from the general result
 in Landau gauge$\cite{ref2}$ and was explicitly calculated
in Ref.$\cite{ref18}$. 
It is obvious that after taking the large-$M$ limit, the amplitude
indeed vanishes. Therefore we can always define the gluon-ghost vertex 
renormalization constant as,
\begin{eqnarray}
Z_2=1.
\label{eq47}
\end{eqnarray} 

\section{Finite Renormalization}

Now let us consider the renormalization of the coupling constant. There are
two ways to implement this: one way is using the Slavnov-Taylor identities
and the known one-loop results to determine the local part of the one-loop
quantum effective action$\cite{ref4,ref6}$; another way, which we shall  
adopt in the following, is to use the 
relations among various coupling constants imposed by the Slavnov-Taylor 
identities to determine the finite renormalization of the coupling 
constant$\cite{ref3,ref19}$.
Since the on-shell renormalization is compatible with the Slavnov-Taylor
identity and the renormalized coupling constant is unique, 
we can write the local effective action in terms of the renormalized fields
in the following two forms,
\begin{eqnarray}
S&=&\int d^3x\left[\frac{1}{2}Z_A\epsilon^{\mu\nu\rho}A_{\mu}^a
\partial_{\nu}A_{\rho}^a+Z_c\bar{c}^a\partial^{\mu}\partial_{\mu}c^a
+Z_{\psi}\bar{\psi}(i/\hspace{-2mm}\partial-Z_mm_{\rm ph})\psi\right.\nonumber\\
&+&\left.\frac{1}{3!}Z_1gf^{abc}\epsilon^{\mu\nu\rho}
A_{\mu}^aA_{\nu}^bA_{\rho}^c-Z_2gf^{abc}\partial^{\mu}\bar{c}^aA_{\mu}^bc^c
+Z_3g\bar{\psi}/\hspace{-2.5mm}A\psi\right]\nonumber\\
&=&\int d^3x\left[\frac{1}{2}Z_A\epsilon^{\mu\nu\rho}A_{\mu}^a
\partial_{\nu}A_{\rho}^a+Z_c\bar{c}^a\partial^{\mu}\partial_{\mu}c^a
+Z_{\psi}\bar{\psi}(i/\hspace{-2mm}\partial-Z_mm_{\rm ph})\psi\right.\nonumber\\
&+&\left.\frac{1}{3!}g_BZ_A^{3/2}\epsilon^{\mu\nu\rho}
A_{\mu}^aA_{\nu}^bA_{\rho}^c-g_BZ_cZ_A^{1/2}
f^{abc}\partial^{\mu}\bar{c}^aA_{\mu}^bc^c
+g_BZ_\psi Z_A^{1/2}\bar{\psi}/\hspace{-2.5mm}A\psi\right],
\label{eq48}
\end{eqnarray}
where the correspondence between the renormalized quantities and
the bare ones is defined as usual,
\begin{eqnarray}
&&A^a_{B\mu}=Z_A^{1/2} A_{\mu}^a,~~ c^a_B=Z_c^{1/2} c^a,~~ 
\bar{c}^a_B=Z_c^{1/2} \bar{c}^a,~~
\psi_B=Z_\psi^{1/2}\psi,\nonumber\\
&&\bar{\psi}_B=Z_\psi^{1/2}\bar{\psi}, ~~m=m_{\rm ph}+\delta m=Z_mm_{\rm ph}.
\label{eq49}
\end{eqnarray}   
Eq.(\ref{eq48}) gives  
\begin{eqnarray}
g=g_BZ_1^{-1}Z_A^{3/2}&=&g_BZ_2^{-1}Z_cZ_A^{1/2}=g_BZ_3^{-1}Z_\psi Z_A^{1/2};
\nonumber\\
\frac{Z_A}{Z_1}&=&\frac{Z_c}{Z_2}=\frac{Z_\psi}{Z_3}.
\label{eq50}
\end{eqnarray}
From Eqs.(\ref{eq26}), (\ref{eq27n}), (\ref{eq46}) and (\ref{eq47}), 
we can see that the relation $Z_c/Z_2=Z_{\psi}/Z_3$
is indeed satisfied. Using Eq.(\ref{eq50}), we obtain 
\begin{eqnarray}
Z_1=1-\frac{g^2}{4\pi}\left(3C_V+\frac{1}{2}\right).
\label{eq51}
\end{eqnarray}

\section{Summary and Discussion}

  We have found the inconsistency of dimensional reduction and
naive dimensional regularization when they are applied to Chern-Simons
type theories. Further we use the consistent dimensional continuation
to re-investigate the one-loop quantum correction of Chern-Simons term 
coupled with spinor fields. As it is pointed out in Ref.$\cite{ref4}$, 
the practice of consistent dimensional regularization requires
the introduction of the higher covariant derivative term like Yang-Mills term,
since this special prescription of dimensional continuation results in the
linear dependence of the $n$-dimensional kinetic operator, even though
the gauge fixing has been performed. Therefore the regularization we adopt
in essence consists of higher covariant derivative regularization combining
with consistent dimensional continuation. 

  With this  regularization prescription, we have calculated all the
one-loop two-point amplitudes and have given
the analytical result of one-loop on-shell quark-gluon vertex with aid of 
 the Slavnov-Taylor identity. In the 
mass-shell renormalization convention, we have found that not only
the coupling constant receives an extra finite 
renormalization from the fermionic loop,
but the fermionic matter also has a finite renormalization. This is different
from the result given in Ref.$\cite{ref3}$, where it was shown that all
the renormalization constants are defined as $Z_i=1$.

 Of course, purely from the viewpoint of renormalization, our results do
not  contradict the ones of Ref.$\cite{ref3}$ since a difference
in a finite renormalization can always be explained as a different choice
of renormalization convention. However, since  Chern-Simons type
theory is finite at one-loop level, the $\beta$-function and the anomalous
dimensions of all the fields vanish identically, and we have no objects like 
renormalization group equation to show the renormalization convention
independence. As pointed out in Ref.$\cite{ref23}$, the only 
criterion for the equivalence among
different renormalization conditions is that all the regularization
schemes preserving the fundamental symmetry such as gauge invariance
should give the same gauge invariant radiative corrections. Therefore,
we interpret these differences as the inconsistency of naive
dimensional regularization. 
 
\acknowledgments
 The financial support of the Academy of Finland under 
the Project No. 37599 is greatly acknowledged. We
would like to thank   
 V.Ya. Fainberg for useful discussions.

\appendix

\section{Slavnov-Taylor Identity from BRST Symmetry}

The BRST transformation reads as follows:
\begin{eqnarray}
\delta A_{\mu}^a&=&-D_{\mu}^{ab}c^b,~\delta c^a=\frac{g}{2}f^{abc}c^bc^c,~
\delta\bar{c}^a=-\frac{1}{\alpha}\partial^{\mu}A_{\mu}^a,\nonumber\\
\delta \psi&=&-igT^ac^a\psi,~\delta\bar{\psi}=ig\bar{\psi}T^ac^a,
\label{eqa1}
\end{eqnarray}
which is nilpotent,
\begin{eqnarray}
\delta^2=0.
\label{eqa2}
\end{eqnarray}
The generating functional with all the external source terms is
\begin{eqnarray}
Z&{\equiv}&Z[J_{\mu}^a,\eta,\bar{\eta},\bar{K},K,\bar{L},L,u^a_{\mu},v^a]
\nonumber\\
&=&\int {\cal D}X\exp\left\{i\left[S+\int d^3x \left(J^{\mu a}A_{\mu}^a+
\bar{\psi}\eta+\bar{\eta}\psi + \bar{K}^ac^a+\bar{c}^aK^a\right)\right.\right.
\nonumber\\
&&\left.\left. +\int d^3x \left(-\bar{L}gT^ac^a\psi+\bar{\psi}gT^ac^aL
-u^a_{\mu}D^{\mu ab}c^b+v^a\frac{g}{2}f^{abc}c^bc^c\right)\right]\right\}.
\label{eqa3}
\end{eqnarray}
The BRST invariance of the generating functional lead to the following
 general Ward identity:
\begin{eqnarray}
&&\delta Z=\int {\cal D}X \left\{\int d^3u \left[J_{\mu}^a \delta A_{\mu}^a
+ \delta \bar{\psi}\eta-\bar{\eta}
\delta \psi+\delta\bar{c}^aK^a-\bar{K}^a\delta c^a\right]\right.\nonumber\\
&& \times \left.\exp\left[iS+i\int (\mbox{the source term})\right]\right\} =0,
\nonumber\\
&& \int d^3 u\left[J_{\mu}^a\frac{\delta}{\delta u^a_{\mu}}-i\bar{\eta}
\frac{\delta}{\delta \bar{L}}-i\frac{\delta}{\delta L}\eta-\bar{K}^a
\frac{\delta}{\delta v^a}-\frac{1}{\alpha}\left(\partial_{\mu}
\frac{\delta}{\delta J^a_{\mu}}\right)K^a\right]Z=0.
\label{eqa4}
\end{eqnarray}
It can be directly written out the Ward identities for the generating
functional of the connected Green functions due to the linearity of
the functional differential operator in Eq.(\ref{eqa4})
\begin{eqnarray}
\int d^3 u\left[J_{\mu}^a\frac{\delta}{\delta u^a_{\mu}}-i\bar{\eta}
\frac{\delta}{\delta \bar{L}}-i\frac{\delta}{\delta L}\eta-\bar{K}^a
\frac{\delta}{\delta v^a}-\frac{1}{\alpha}\left(\partial_{\mu}
\frac{\delta}{\delta J^a_{\mu}}\right)K^a\right]W=0,
\label{eqa5}
\end{eqnarray}
where $Z=\exp[iW]$. Acting $\delta/\delta \bar{\eta}(x)$,
$\delta/\delta {\eta}(y)$ and $\delta/\delta K^a(z)$ on above identity
and then the external sources to zero, we obtain the Ward identity
containing the quark-gluon vertex,
\begin{eqnarray}
&&\left[\frac{1}{\alpha}\frac{\delta}{\delta \bar{\eta}(x)}
\frac{\delta}{\delta {\eta}(y)}\partial_{\mu}\frac{\delta}{\delta J^a_{\mu}(z)}
+i\frac{\delta}{\delta {\eta}(y)}\frac{\delta}{\delta K^a(z)}
\frac{\delta}{\delta \bar{L}(x)}\right.\nonumber\\
&& \left.+i\frac{\delta}{\delta \bar{\eta}(x)}\frac{\delta}{\delta K^a(z)}
\frac{\delta}{\delta L(y)}\right]W|_{\mbox{\rm all the external sources $=0$}}
=0,
\nonumber\\
&&\frac{1}{\alpha}\frac{\partial}{\partial z_{\mu}}\langle\psi(x)\bar{\psi}(y)
A_{\mu}^a(z)\rangle_C
 +igT^b\langle\bar{\psi}(y)\bar{c}^a(z)c^b(x)\psi (x)\rangle_C\nonumber\\
&&-igT^b\langle\psi(x)\bar{c}^a(z)\bar{\psi}(y)c^b(y)\rangle_C =0,
\label{eqa6}
\end{eqnarray}
where the subscript `` C '' means the connected part of the Green functions.
Decomposing the above Green functions into 1PI part, we get 
\begin{eqnarray}
&&\frac{1}{\alpha}\frac{\partial}{\partial z_{\mu}}
\int d^3u d^3v d^3 w \,iD^{aa'}_{\mu\nu}(z-w) iS(x-u) g \Gamma^{a'}(u,v,w)
iS(v-y)\nonumber\\
&&+ig\int d^3u d^3v\left[\gamma^{a'}(x,u,v)iS(u-y)
-iS(x-u)\gamma^{a'}(u,y,v)\right] iD^{a'a}(v-z)=0,
\label{eqa7}
\end{eqnarray}
where $\Gamma^{a}(u,v,w)$ is the 1PI part of the fermion-gluon vertex function, 
$\gamma^{a}(x,u,v)$ and $\gamma^{a}(u,y,v)$ are the composite ghost-gluon
vertex functions.
After Fourier transformation, we obtain 
\begin{eqnarray}
&&\frac{1}{\alpha}r^{\mu}D^{aa'}_{\mu\nu}(r)S(p)\Gamma^{\nu a'}(p,q,r)S(q)
+\gamma^{a'}(p,q,r)D^{aa'}(r)S(q)\nonumber\\
&&-S(p)\gamma^{a'}(p,q,r) D^{a'a}(r)=0,
\label{eqa8}
\end{eqnarray}
where $r_{\mu}=q_{\mu}-p_{\mu}$.
Considering the fact that the longitudinal part of gauge field receives no 
quantum correction, i.e. 
\begin{eqnarray}
r^{\mu}D_{\mu\nu}^{ab}(r)
=r^{\mu}D_{\mu\nu}^{ab(0)}(r)=-\alpha \frac{r_{\nu}}{r^2}
\delta^{ab},
\label{eqa9} 
\end{eqnarray}
and using the general form of the full ghost propagator
\begin{eqnarray}
D^{ab}(r)=-\frac{i\delta^{ab}}{r^2\left[1+\Sigma_g(r^2)\right]},
\label{eqa10}
\end{eqnarray}
we  get the required Ward identity
\begin{eqnarray}
r^{\mu}\Gamma^a_{\mu}(p,q,r)\left[1+\Sigma_g(r^2)\right]
=\gamma^{a}(p,q,r)S^{-1}(q)-S^{-1}(p)\gamma^{a}(p,q,r).
\label{eqa11}
\end{eqnarray} 
Expanding the above identity up to one-loop order and using
\begin{eqnarray}
&& \Gamma^a_{\mu}(p,q,r)=\gamma_{\mu}T^a+g^2\Gamma_{\mu}^{(1)a}+{\cal O}(g^4),~~
\Sigma_g(r^2)=g^2\Sigma_g^{(1)}(r^2)+{\cal O}(g^4),~~\nonumber\\
&& S^{-1}(p)=/\hspace{-2mm}p-m-g^2 \Sigma^{(1)}(p)+{\cal O}(g^4) ,~~
\gamma^a(p,q,r)=T^a+g^2\gamma^{(1)a}(p,q,r)+{\cal O}(g^4),
\label{eqa12}
\end{eqnarray}
we obtain the desired one-loop Slavnov-Taylor identity:  
\begin{eqnarray}
(q^{\mu}-p^{\mu})\Gamma_{\mu}^{(1)a}(p,q,r)&=&
-(/\hspace{-2mm}q-/\hspace{-2mm}p)T^a
\Sigma_g^{(1)}(r^2)-T^a\left[\Sigma^{(1)}(q)-\Sigma^{(1)}(p)\right]
\nonumber\\
&+&g^2\left[\gamma^{(1)a}(p,q,r)(/\hspace{-2mm}q-m)-
(/\hspace{-2mm}p-m)\gamma^{(1)a}(p,q,r)\right].
\label{eqa13}
\end{eqnarray}

\begin{figure}
\centering
\input FEYNMAN
\begin{picture}(25000,8000)
\THICKLINES
\put(14000,5000){\circle{3000}}
\startphantom
\drawline\gluon[\E\REG](0,0)[2]\gluoncap
\stopphantom
\pbackx=12600 \pbacky=5000
\global\multiply\plengthx by -1
\global\multiply\plengthy by -1
\global\advance\pbackx by \plengthx
\global\advance\pbacky by \plengthy
\drawline\gluon[\E\REG](\pbackx,\pbacky)[2]\gluoncap
\drawline\gluon[\W\FLIPPED](\gluonfrontx,\gluonfronty)[2]\gluoncap
\gluonbackx=15400 \gluonbacky=5000
\negate\gluonlengthx
\negate\gluonlengthy
\global\advance\gluonbackx by \gluonlengthx
\global\advance\gluonbacky by \gluonlengthy
\drawline\gluon[\W\FLIPPED](\gluonbackx,\gluonbacky)[2]\gluoncap
\drawline\gluon[\E\REG](\gluonfrontx,\gluonfronty)[2]\gluoncap
\end{picture}
\caption{\protect\small Contribution to vacuum polarization from 
fermionic loop}

\begin{picture}(8000,8000)
\THICKLINES
\drawline\fermion[\E\REG](0,0)[2000]
\drawloop\gluon[\N 5](\pbackx,\pbacky)
\drawline\fermion[\E\REG](\pbackx,\pbacky)[2000]
\drawline\fermion[\W\REG](\pbackx,\pbacky)[7000]
\end{picture}
\caption{\protect\small Self-energy for fermionic field }

\vspace{5mm}

\begin{picture}(10000,5000)
\THICKLINES
\drawline\scalar[\E\REG](0,0)[2]
\drawloop\gluon[\NE 3](\pbackx,\pbacky)
\drawline\scalar[\E\REG](\pbackx,\pbacky)[2]
\drawline\scalar[\W\REG](\pbackx,\pbacky)[4]
\end{picture}
\caption{\protect\small Self-energy for ghost field }

\vspace{5mm}

\begin{picture}(40000,20000)
\THICKLINES
\drawline\gluon[\S\REG](9000,11000)[4]
\drawline\fermion[\SW\REG](\gluonbackx,\gluonbacky)[6400]
\drawline\fermion[\W\REG](\fermionbackx,\fermionbacky)[3000]
\drawarrow[\E\ATBASE](\pmidx,\pmidy)
\put(2000,3000){$p$}
\drawline\fermion[\SE\REG](\gluonbackx,\gluonbacky)[6400]
\drawline\fermion[\E\REG](\fermionbackx,\fermionbacky)[3000]
\drawarrow[\W\ATBASE](\pmidx,\pmidy)
\put(15000,3000){$q$}
\drawline\gluon[\W\REG](\fermionfrontx,\fermionfronty)[8]
\put(8000,200){$(a)$}

\drawvertex\gluon[\S 3](31000,11500)[4]
\drawline\fermion[\W\REG](\vertexthreex,\vertexthreey)[3000]
\drawarrow[\E\ATBASE](\pmidx,\pmidy)
\put(24000,3000){$p$}
\drawline\fermion[\E\REG](\vertexthreex,\vertexthreey)[9600]
\drawline\fermion[\E\REG](\vertextwox,\vertextwoy)[3000]
\drawarrow[\W\ATBASE](\pmidx,\pmidy)
\put(37000,3000){$q$}
\put(30000,200){$(b)$}
\end{picture}
\caption{\protect\small One-loop quark-gluon vertex}

\begin{picture}(40000,20000)
\THICKLINES
\drawline\fermion[\SW\REG](9000,10000)[6000]
\put(8600,10100){$\bigotimes$}
\drawline\scalar[\SE\REG](\fermionfrontx,\fermionfronty)[3]
\drawline\fermion[\W\REG](\fermionbackx,\fermionbacky)[3000]
\drawarrow[\E\ATBASE](\pmidx,\pmidy)
\put(2000,5000){$p$}
\drawline\gluon[\W\REG](\scalarbackx,\scalarbacky)[8]
\drawline\scalar[\E\REG](\gluonfrontx,\gluonfronty)[2]
\drawarrow[\W\ATBASE](\pmidx,\pmidy)
\put(15000,5000){$r=q-p$}
\put(8000,2000){$\gamma^a(p,q,r)$}

\drawline\scalar[\SW\REG](32000,10000)[3]
\put(31600,10100){$\bigotimes$}
\drawline\fermion[\SE\REG](\scalarfrontx,\scalarfronty)[6000]
\drawline\scalar[\W\REG](\scalarbackx,\scalarbacky)[2]
\drawarrow[\E\ATBASE](\pmidx,\pmidy)
\put(23000,5000){$r=q-p$}
\drawline\gluon[\W\REG](\fermionbackx,\fermionbacky)[8]
\drawline\fermion[\E\REG](\gluonfrontx,\gluonfronty)[3000]
\drawarrow[\W\ATBASE](\pmidx,\pmidy)
\put(38000,5000){$q$}
\put(30000,2000){$\gamma^a(p,q,r)$}
\end{picture}
\caption{\protect\small One-loop ghost-fermion composite vertex}

\begin{picture}(40000,15000)
\THICKLINES
\drawline\gluon[\S\REG](9000,11000)[3]
\drawline\scalar[\SW\REG](\gluonbackx,\gluonbacky)[3]
\drawline\scalar[\W\REG](\scalarbackx,\scalarbacky)[2]
\drawline\scalar[\SE\REG](\gluonbackx,\gluonbacky)[3]
\drawline\scalar[\E\REG](\scalarbackx,\scalarbacky)[2]
\drawline\gluon[\W\REG](\scalarfrontx,\scalarfronty)[8]
\put(8000,200){$(a)$}

\drawvertex\gluon[\S 3](31000,11200)[3]
\drawline\scalar[\W\REG](\vertexthreex,\vertexthreey)[2]
\drawline\scalar[\E\REG](\vertexthreex,\vertexthreey)[3]
\drawline\scalar[\E\REG](\scalarbackx,\scalarbacky)[2]
\put(30000,200){$(b)$}
\end{picture}
\caption{\protect\small  One-loop ghost-gluon vertex}

\end{figure}
\end{document}